\documentclass[twocolumn,showpacs,aps,floatfix,prd,nofootinbib]{revtex4}
\usepackage{graphics}
\usepackage{graphicx}
\usepackage{amssymb,amsmath}
\usepackage{relsize}

\def\babar{\mbox{\slshape B\kern-0.1em{\smaller A}\kern-0.1em
    B\kern-0.1em{\smaller A\kern-0.2em R}}}
\newcommand{\BABARPubNumber}{19/003}
\newcommand{\SLACPubNumber}{17492}

\begin{document}

\begin{minipage}{.45\linewidth}
\begin{flushleft}
\babar\--PUB-\BABARPubNumber \\
SLAC-PUB-\SLACPubNumber \\
\end{flushleft}
\end{minipage}

\title{\bf\boldmath
Resonances in $e^+e^-$ annihilation near 2.2~GeV}

\begin{abstract}
Using the initial-state radiation method, the $e^+e^- \to K_S K_L$ cross 
section from 1.98 to 2.54 GeV is measured in a data sample of 469 fb$^{-1}$
collected with the \babar\ detector. The results are used in conjunction with 
previous \babar\ results for the $e^+e^- \to K^+ K^-$, $e^+e^- \to \pi^+\pi^-$,
$e^+e^- \to \pi^+\pi^-\eta$, and $e^+e^- \to \omega\pi\pi$ cross sections
to investigate the nature of the resonance structure recently observed by
the BESIII experiment in the $e^+e^- \to K^+ K^-$ cross section.
\end{abstract}

\author{J.~P.~Lees}
\author{V.~Poireau}
\author{V.~Tisserand}
\affiliation{Laboratoire d'Annecy-le-Vieux de Physique des Particules (LAPP), Universit\'e de Savoie, CNRS/IN2P3,  F-74941 Annecy-Le-Vieux, France}
\author{E.~Grauges}
\affiliation{Universitat de Barcelona, Facultat de Fisica, Departament ECM, E-08028 Barcelona, Spain }
\author{A.~Palano}
\affiliation{INFN Sezione di Bari and Dipartimento di Fisica, Universit\`a di Bari, I-70126 Bari, Italy }
\author{G.~Eigen}
\affiliation{University of Bergen, Institute of Physics, N-5007 Bergen, Norway }
\author{D.~N.~Brown}
\author{Yu.~G.~Kolomensky}
\affiliation{Lawrence Berkeley National Laboratory and University of California, Berkeley, California 94720, USA }
\author{M.~Fritsch}
\author{H.~Koch}
\author{T.~Schroeder}
\affiliation{Ruhr Universit\"at Bochum, Institut f\"ur Experimentalphysik 1, D-44780 Bochum, Germany }
% ...
\collaboration{The \babar\ Collaboration}
\noaffiliation

\maketitle

\section{Introduction\label{intro}}
Recently, a precise measurement of the $e^+e^-\to K^+K^-$ cross 
section in the center-of-mass (c.\ m.) energy range $E=$2.00--3.08~GeV was
performed by the BESIII Collaboration~\cite{beskkc}. In this cross section,
a clear interference pattern was observed near 2.2~GeV. To explain this 
pattern, BESIII inferred the existence of a resonance with a mass
of $2239\pm7\pm11$~MeV/$c^2$ and a width of $140\pm12\pm21$~MeV.
In the Particle Data Group (PDG) table~\cite{pdg} there are two vector 
resonances with a mass near 2.2~GeV/$c^2$: $\phi(2170)$ and $\rho(2150)$.
The first is observed in three reactions: 
$e^+e^-\to \phi(2170)$~\cite{babar2170,belle2170},
$J/\psi\to \eta\phi(2170)$~\cite{bes2170,besiii2170}, and
$e^+e^-\to \eta\phi(2170)$~\cite{besiii2170n}, but only in the decay mode 
$\phi(2170) \to \phi(1020) f_0(980)$. As shown in Ref.~\cite{beskkc},
the parameters of the resonance structure observed in the $e^+e^-\to K^+K^-$
cross section differ from the $\phi(2170)$ PDG parameters 
by  more  than  3$\sigma$ in  mass  and  more than 2$\sigma$ in width.
The isovector resonance $\rho(2150)$ is not well established.
The PDG lists three $e^+e^-$ annihilation processes in which 
evidence for its existence is seen: $e^+e^-\to f_1(1275)\pi^+\pi^-$,
$e^+e^-\to \eta^\prime\pi^+\pi^-$, and $e^+e^-\to \pi^+\pi^-$. In the
first two reactions, wide ($\Gamma\sim 300$~MeV) resonance-like structures
are observed near the reaction thresholds~\cite{ompipi1}. 
A completely different structure is seen in the third process.
A resonance with mass and width $2254\pm22$~MeV/$c^2$ and $109\pm76$~MeV, 
respectively, is needed to describe the interference pattern in the
$e^+e^-\to \pi^+\pi^-$ cross section~\cite{pipi}.
Note that the parameters of this resonance are very similar to those
mentioned above for the $e^+e^-\to K^+ K^-$ reaction from BESIII.

Any resonance in the $e^+e^-\to K^+ K^-$ cross section should also be present
in $e^+e^-\to K_S K_L$. The most precise data on this reaction near 2~GeV were
obtained by the \babar\ Collaboration~\cite{kskl}. In this previous work,
the $e^+e^-\to K_S K_L$ cross section was measured up to 2.2~GeV. 
Above 2~GeV, the cross section was found to be consistent with zero within
the statistical uncertainties of around 20 pb.
In the present work we expand the energy region of the \babar\ $K_S K_L$ 
measurement up to 2.5~GeV. 
The new $K_S K_L$ measurements, in conjunction with previous \babar\ results
for other exclusive $e^+e^-$ processes, are used to investigate
the nature of the structure observed by BESIII in $e^+e^-\to K^+ K^-$.

\section{\bf\boldmath FIT to the BESIII and \babar\ $e^+e^-\to K^+ K^-$
data \label{fitkk}}
\begin{figure}
\includegraphics[width=0.45\textwidth]{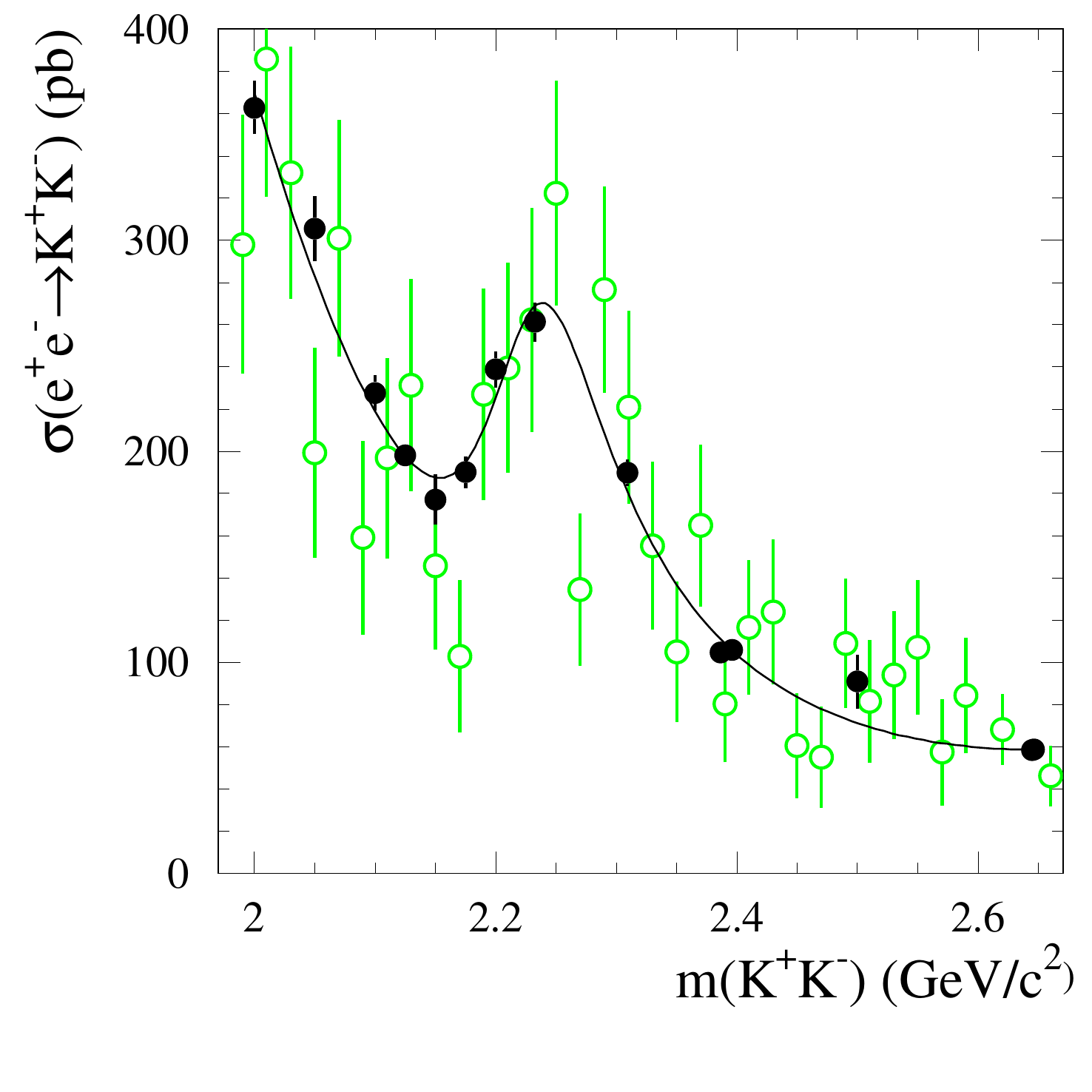}
\caption{The $e^+e^-\to K^+K^-$ cross section measured by 
BESIII~\cite{beskkc} (filled circles) and \babar~\cite{babarkkc} (open circles).
The curve is the result of the fit to a coherent sum of resonant and 
nonresonant contributions (see text).
\label{fig1}}
\end{figure}
In Fig.~\ref{fig1} we show BESIII~\cite{beskkc} and \babar\ \cite{babarkkc}
data on the dressed Born cross
section for the process $e^+e^-\to K^+K^-$ in the energy region of interest. 
The dressed cross section used to obtain resonance parameters is calculated 
from the bare cross section ($\sigma_{\rm b}$) listed in 
Refs.~\cite{beskkc, babarkkc} as
 $\sigma=\sigma_{\rm b} R_{\rm VP}/C_{\rm FS}$, where $R_{\rm VP}$ is 
the factor taking into account the vacuum polarization correction, while 
$C_{\rm FS}$ is the final-state correction (see, e.g., Ref.~\cite{fscor}). 
The latter, in particular, takes into account extra photon radiation from the 
final state. In the energy region of interest,
2.00--2.5~GeV, $R_{\rm VP}\approx1.04$ and $C_{\rm FS}=1.008$. The 
BESIII and \babar\ data on the dressed $e^+e^-\to K^+K^-$ cross
section are fitted by a coherent sum of resonant and nonresonant 
contributions 
\begin{equation} 
\sigma(E)=\frac{M_{R}^2\beta(E)^3}{E^2\beta(M_{R})^3}
\left |\sqrt{\sigma_{R}}BW(E)+e^{i\varphi}P(E) \right |^2,
\label{signaleq}
\end{equation}
where $\beta(E)=\sqrt{1-4m_K^2/E^2}$, $m_K$ is the charged kaon mass, 
$BW(E)=M_{R}\Gamma_{R}/(M_{R}^2-E^2-iE\Gamma_{R})$ is the Breit-Wigner
function describing the resonant amplitude, $M_R$, $\Gamma_R$, and 
$\sigma_{R}$ are the resonance mass, width, and peak cross section,
$P(E)$ is a second-order polynomial describing the nonresonant amplitude, 
and $\varphi$ is the relative phase between the resonant and nonresonant
amplitudes. The fit result is shown in Fig.~\ref{fig1}.
The fit yields $\chi^2/\nu=55.8/40$ ($P(\chi^2)=5\%$)
and the fit parameters are listed in Table~\ref{tab0}.
\begin{table}
\caption{The parameters for the fit to 
the $e^+e^-\to K^+K^-$ cross section data from BESIII and \babar . The quoted uncertainties are
statistical and systematic, respectively.\label{tab0}}
\begin{ruledtabular}
\begin{tabular}{cc}
$M_R$                   & $2227\pm 9\pm 9$~MeV/$c^2$\\
$\Gamma_{R}$            & $127\pm 14\pm 4$~MeV\\
$\sigma_R$              & $39 \pm 6\pm 4$~pb\\
$\varphi$              & $143 \pm 8\pm 9$~deg\\
\end{tabular}
\end{ruledtabular}
\end{table}

The systematic uncertainties in the resonance parameters come mainly from
uncertainties in the description of the resonance and nonresonance shapes. 
The uncertainty due to the absolute c.\ m.\ energy calibration is 
negligible~\cite{beskkc,babarkkc}. For the signal shape we study the
effect of the energy-dependent width assuming that the main resonance decay
mode is either $K^+K^-$ or $\eta\rho$. We also use another parametrization
of the nonresonance amplitude, in which the main energy dependence is given by
the function $a/(E^2-b^2)$ inspired by the vector-meson dominance model,
where $a$ and $b$ are fitted parameters,
while small deviations from the main dependence are described by a quadratic 
polynomial. The nonresonance amplitude may have an energy-dependent
imaginary part originating from vector resonances lying below 2~GeV.
Using the results of Ref.~\cite{bel}, we estimate that
its fraction reaches 10\% at 2~GeV and decreases to 5\% at 2.5~GeV. 
To study the effect of the imaginary parts, we multiply the function $P(E)$ in 
Eq.~(\ref{signaleq}) by a factor of $1\pm iG(E)$, where $G(E)$ is a linear 
function decreasing from 0.05--0.15 at $E=2$~GeV to zero at 2.5~GeV. The 
deviations from the nominal parameter values listed in Table~\ref{tab0} are 
taken as the estimates of the systematic uncertainties given in Table~\ref{tab0}.
The systematic uncertainty in the parameter $\sigma_R$ 
includes also the correlated systematic uncertainty in the $e^+e^-\to K^+K^-$
cross section, which is 2.5\% (6\%) for the BESIII (\babar) data. 

Our values for the resonance mass and width are close to the values
$2239\pm7\pm11$~MeV/$c^2$ and $140\pm12\pm21$~MeV obtained in Ref.~\cite{beskkc}.
We also perform the fit to the \babar\ data only. The resulting parameters
are $M_R=2201\pm19$~MeV/$c^2$, $\Gamma_R=70\pm 38$~MeV, and 
$\sigma_R=42^{+29}_{-16}$ pb. The resonance significance in the \babar\ data
estimated from the $\chi^2$ difference for the fits with and without the 
resonance contribution is $3.5\sigma$.

\section{\bf\boldmath The $e^+e^-\to K_S K_L$ cross section in the 
2.0--2.5~GeV energy range \label{kskl}}
The data analysis presented in this paper is based on methods developed for the
measurement of the $e^+e^-\to K_S K_L$ cross section in Ref.~\cite{kskl}. The
data set, with an integrated luminosity of 469 fb$^{-1}$~\cite{lum}, was 
collected with the \babar\ detector~\cite{BABAR} at the SLAC PEP-II
asymmetric-energy $e^+e^-$ storage ring at the $\Upsilon(4S)$ 
resonance and 40~MeV below this resonance. The initial-state-radiation (ISR)
technique is used, in which the cross section for the process 
$e^+e^-\to K_S K_L$ is determined from the $K_S K_L$ invariant mass spectrum
measured in the reaction $e^+e^-\to K_S K_L\gamma$. 

The selection criteria for $e^+e^-\to K_S K_L\gamma$ events 
are described in detail in Ref.~\cite{kskl}. We require the detection of 
all the final-state particles. The ISR photon candidate must have an energy in
the c.\ m.\ frame greater than 3~GeV. The $K_S$ candidate is reconstructed
using the $K_S\to \pi^+\pi^-$ decay mode. Two oppositely charged tracks not 
identified as electrons are fitted to a common vertex. The distance between 
the reconstructed  $K_S$ decay vertex and the beam axis must be in the range from 0.2 to
40.0 cm. The cosine of the angle between a vector from the beam interaction 
point to the 
$K_S$ vertex and the $K_S$ momentum in the plane transverse to the beam axis
is required to be larger than 0.9992. The invariant mass of the $K_S$
candidate must be in the range 0.482--0.512~GeV/$c^2$. 
The $K_L$ candidate is a cluster in the calorimeter with energy deposition
greater than 0.2~GeV. 
To suppress background, we also require the event to not contain 
extra charged tracks originating from the interaction region or extra 
photons with energy larger than 0.5~GeV.

The ISR photon, $K_S$, and $K_L$ candidates are subjected to a three-constraint
kinematic fit to the $e^+e^-\to K_S K_L\gamma$ hypothesis with the requirement
of energy and momentum balance. Only the angular information is used in the fit
for the $K_L$ candidate. If there are several $K_L$ candidates in an event,
the $K_S K_L\gamma$ combination giving the smallest $\chi^2$ value is
retained. The particle parameters after the kinematic fit are used to 
calculate the $K_S K_L$ invariant mass $m(K_S K_L)$, which is required to 
satisfy $1.06 < m(K_S K_L) < 2.5$~GeV/$c^2$.
\begin{figure}
\includegraphics[width=0.45\textwidth]{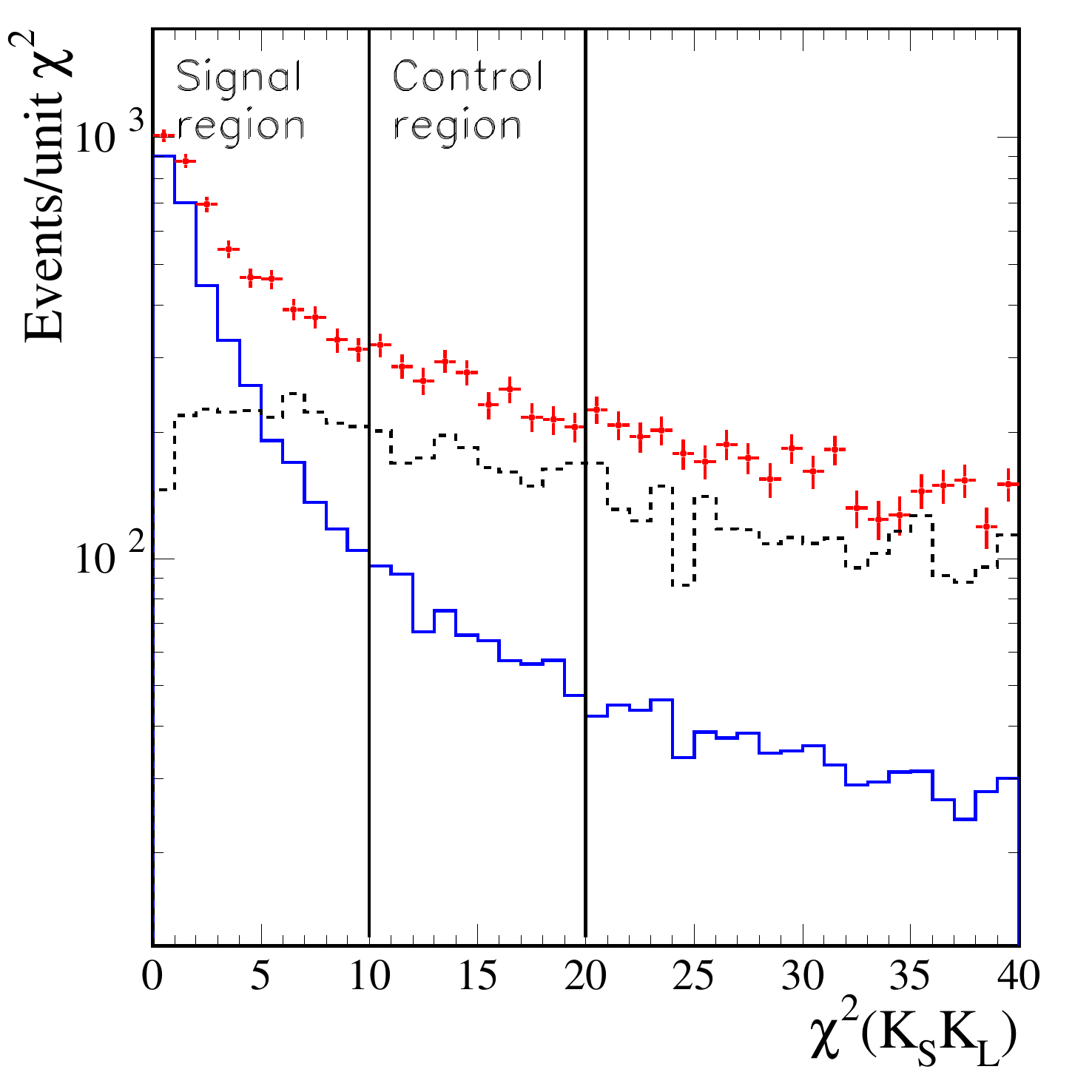}
\caption{The kinematic-fit $\chi^2$ distribution for selected data events with
$1.06 < m(K_S K_L) < 2.5$~GeV/$c^2$ (points with error bars). The hatched 
histogram represents the simulated background contribution.
The solid histogram shows the simulated signal distribution. The vertical
lines indicate the boundaries of the signal and control regions.
\label{fig2}}
\end{figure}

The $\chi^2$ distribution from the fit for the selected events is shown in 
Fig.~\ref{fig2} in comparison
with the simulated signal and background distributions. The background is 
dominated by the ISR processes $e^+e^-\to K_SK_L\pi^0\gamma$,
$K_SK_L\eta\gamma$, and $K_SK_L\pi^0\pi^0\gamma$. The condition
$\chi^2 < 10$ is applied to select signal events. The control region 
$10<\chi^2 < 20$ is used to estimate and subtract background. The numbers of 
signal ($N_s$) and background ($N_b$) events in the signal region 
($\chi^2 < 10$) are determined as
\begin{equation} 
N_s=N_1-N_b,\;
N_b=(N_2-aN_s)/b,\label{bkgsub}
\end{equation}
where $N_1$ and $N_2$ are the numbers of selected data events in the signal
and control regions, and $a=0.20\pm0.01$ and $b=0.87\pm 0.09$ are 
the $N_2/N_1$ ratios for signal and background, respectively.

The value of the coefficient $a$ is determined from the simulated signal $\chi^2$
distribution. For the mass region of interest 
$2.0 < m(K_S K_L) < 2.5$~GeV/$c^2$ , where
the number of signal events is small, the $aN_s$ term in the expression for 
$N_b$ is negligible. The coefficient $b$ is determined in two ways: either 
using background simulation, or from the difference between the data and simulated
signal distributions in Fig.~\ref{fig2}. The signal distribution is
normalized to the number of data events with $\chi^2 < 3$ after subtraction
of the background estimated from simulation. The average of the two $b$ values
is quoted above. Their difference (10\%) is taken as an estimate of the
systematic uncertainty in $b$. As shown in Ref.~\cite{kskl}, the
background $m(K_S K_L)$ distribution obtained using Eq.~(\ref{bkgsub})
is found to be in reasonable agreement with the same distribution obtained 
from simulation.
\begin{figure}
\includegraphics[width=0.45\textwidth]{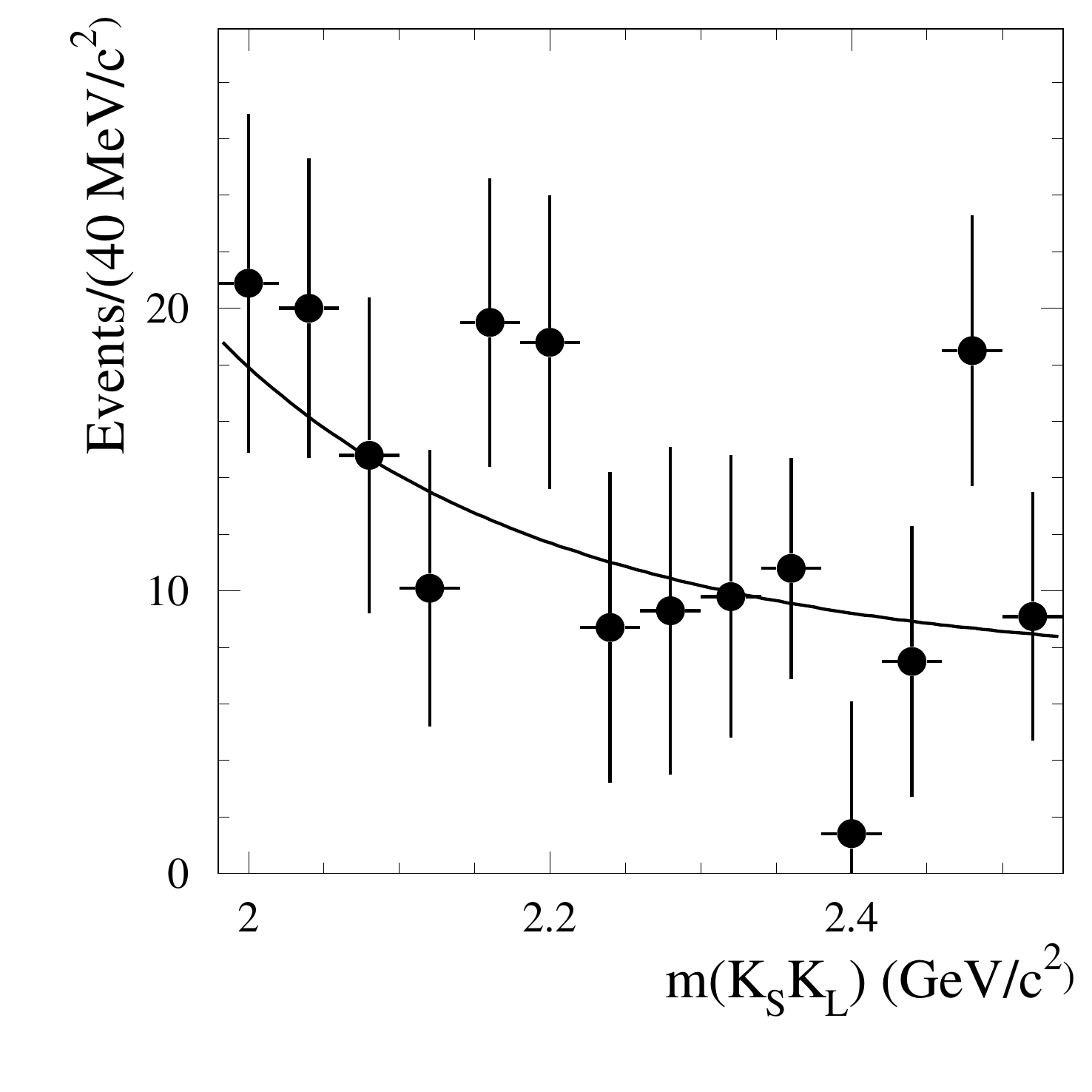}
\caption{
The $m(K_S K_L)$ distribution for data events with $\chi^2 < 10$.
The curve represents background estimated from the control region. 
\label{fig3}}
\end{figure}

The background 
estimated from the control region
decreases monotonically with
increasing $m(K_S K_L)$ and is well approximated by a smooth function. 
Figure~\ref{fig3} shows the $m(K_S K_L)$ distribution for data events from
the signal region. The curve represents the estimated background distribution.

\begin{figure*}
\includegraphics[width=0.45\textwidth]{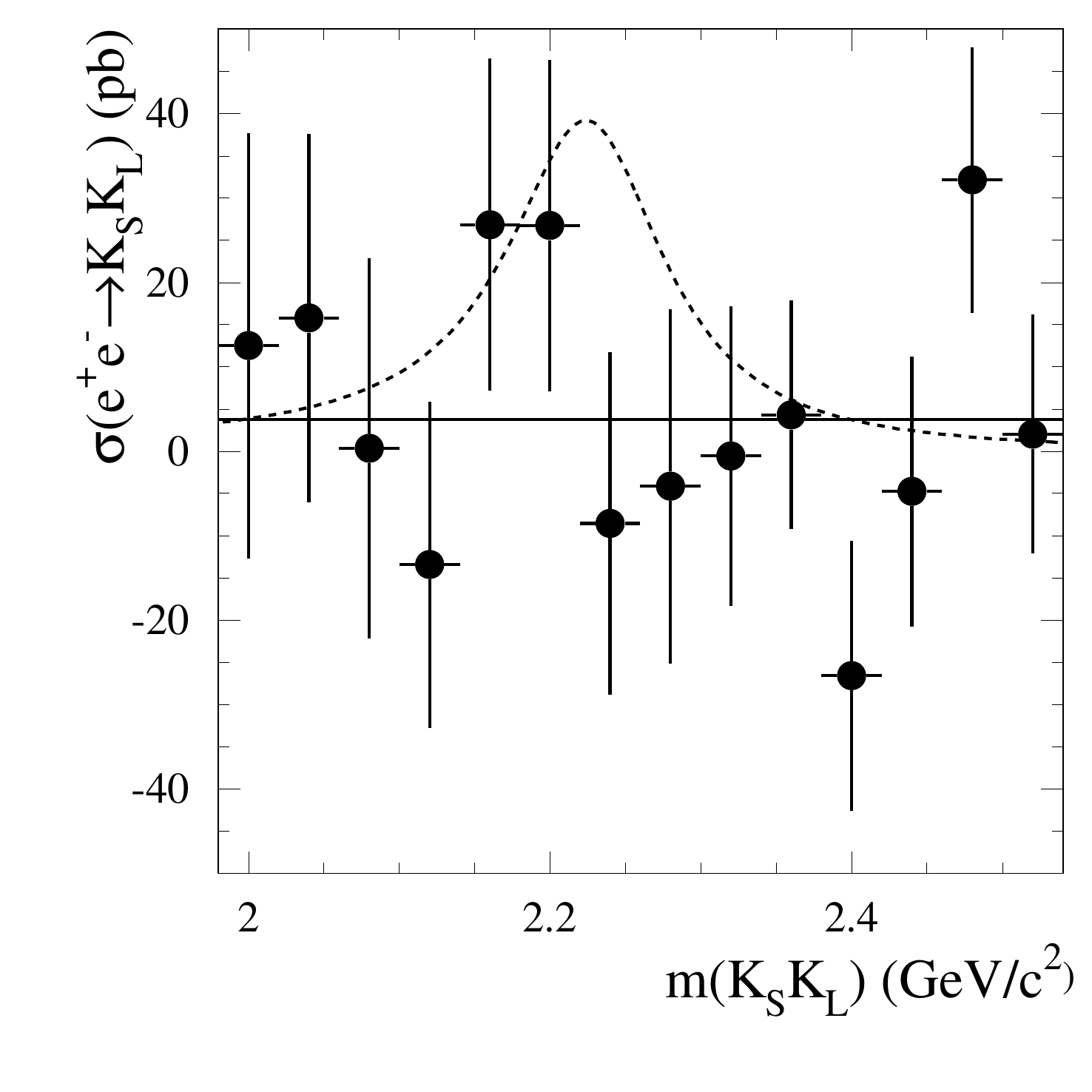}
\includegraphics[width=0.45\textwidth]{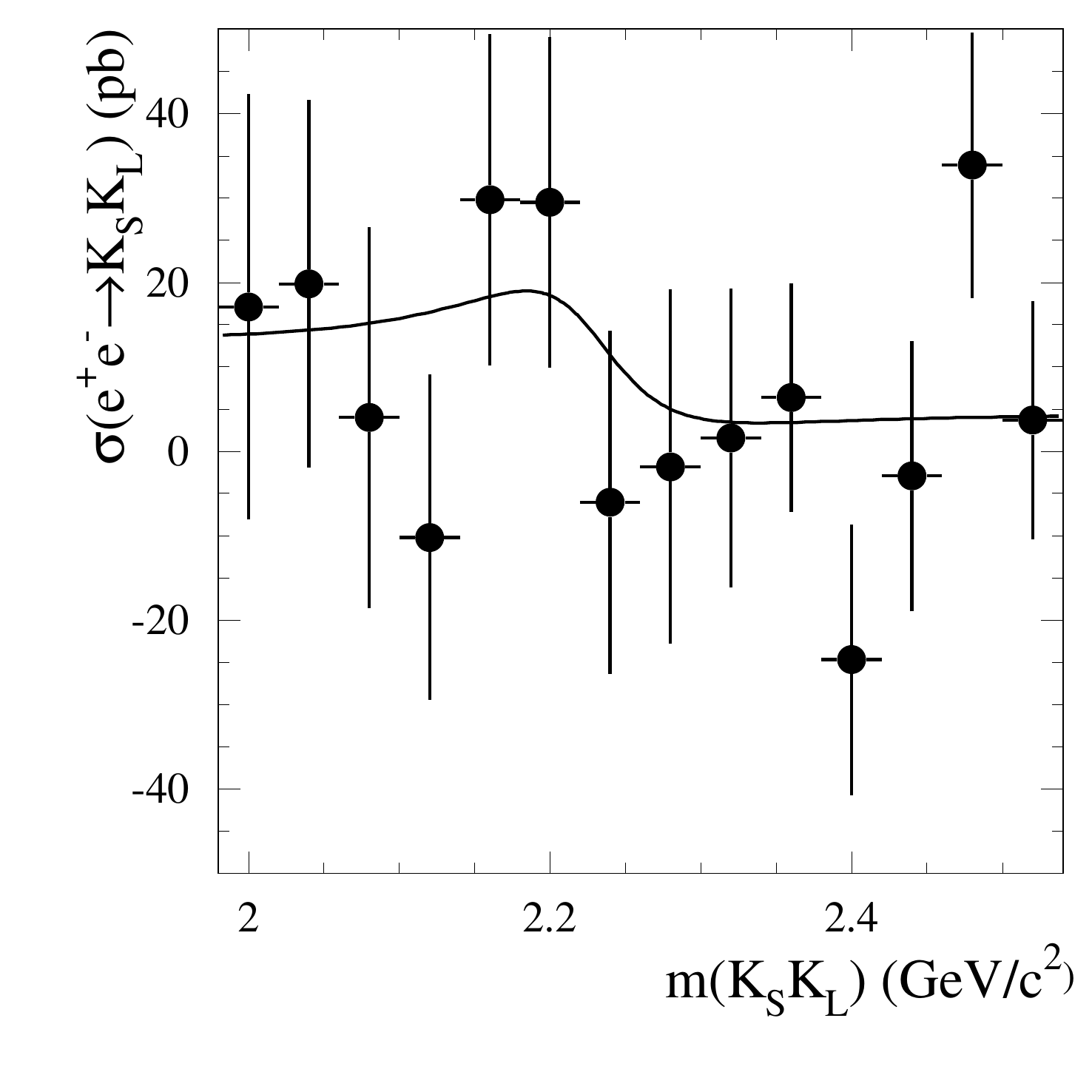}
\caption{Left panel: The measured $e^+e^-\to K_S K_L$ cross section fitted
with a constant (solid line). The dashed curve represents the resonance
line shape with the parameters listed in Table~\ref{tab0}.
Right panel: The curve is the result of the fit to the $e^+e^-\to K_S K_L$ 
data with a coherent sum of a resonant amplitude with the parameters
listed in Table~\ref{tab0}
and a nonresonant constant amplitude. The points with error bars represent
the data following subtraction of the background, which has been scaled by a
factor of 0.94 (see text).
\label{fig4}}
\end{figure*}

The uncertainty in the background is 12\%, which includes the 10\% uncertainty
in the parameter $b$ in Eq.~(\ref{bkgsub}) and a 6\% uncertainty
in the background approximation.
We do not see a significant signal of $K_S K_L$ events over 
background. The $e^+e^-\to K_S K_L$ cross section in the mass region 
$1.96<m(K_S K_L)<2.56$~GeV/$c^2$ obtained from the mass spectrum 
in Fig.~\ref{fig3} after background subtraction is shown in 
Fig.~\ref{fig4} (left). The details on the 
detection efficiency and ISR luminosity can be found in Ref.~\cite{kskl}. The
numerical values of the $e^+e^-\to K_S K_L$ cross section, with statistical
and systematic uncertainties, are listed in Table~\ref{tab1}.
\begin{table*}
\caption{The $m(K_S K_L)$ interval and measured 
Born cross sections for the processes $e^+e^-\to K_S K_L$. The quoted
uncertainties are statistical and systematic, respectively.\label{tab1}}
\begin{ruledtabular}
\begin{tabular}{cccc}
$m(K_S K_L)$ (GeV/$c^2$) & $\sigma$ (pb) & $m(K_S K_L)$ (GeV/$c^2$) & $\sigma$ (pb) \\
\hline
1.98--2.02 & $ 12.5 \pm 25.2 \pm  9.2 $ & 2.26--2.30 & $  -4.1 \pm 21.0 \pm  4.6 $ \\
2.02--2.06 & $ 15.8 \pm 21.8 \pm  8.1 $ & 2.30--2.34 & $  -0.6 \pm 17.7 \pm  4.3 $ \\
2.06--2.10 & $  0.4 \pm 22.5 \pm  7.2 $ & 2.34--2.38 & $   4.3 \pm 13.5 \pm  4.0 $ \\
2.10--2.14 & $-13.4 \pm 19.3 \pm  6.5 $ & 2.28--2.42 & $ -26.6 \pm 16.0 \pm  3.8 $ \\
2.14--2.18 & $ 26.9 \pm 19.6 \pm  5.9 $ & 2.42--2.46 & $  -4.8 \pm 16.0 \pm  3.6 $ \\
2.18--2.22 & $ 26.8 \pm 19.6 \pm  5.4 $ & 2.46--2.50 & $  32.1 \pm 15.7 \pm  3.5 $ \\
2.22--2.26 & $ -8.6 \pm 20.3 \pm  5.0 $ & 2.50--2.54 & $   2.0 \pm 14.1 \pm  3.3 $ \\
\end{tabular}
\end{ruledtabular}
\end{table*}
The systematic uncertainties arise mainly from the background subtraction
and are fully correlated between different $m(K_S K_L)$ intervals. 

A fit to the cross section data with a constant yields $\chi^2/\nu=11.7/13$,
where $\nu$ is the number of degrees of freedom. The average value of the
$e^+e^-\to K_S K_L$ cross section between 1.98 and 2.54~GeV/$c^2$
is found to be ($4 \pm 5 \pm 5$)~pb, which is therefore consistent with zero.
The dashed curve in
Fig.~\ref{fig4} (left) represents the cross section for the resonance
with the parameters listed in Table~\ref{tab0}. Formally, from the $\chi^2$ difference
between the two hypotheses in Fig.~\ref{fig4} (left) the resonance 
interpretation can be excluded at $2.3\sigma$. However, possible destructive 
interference between the resonant and nonresonant $e^+e^-\to K_S K_L$ 
amplitudes may significantly weaken this constraint. We also must take 
into account the uncertainty in the background subtraction and the statistical
uncertainty in the resonance cross section obtained from the fit to
the $e^+e^-\to K^+ K^-$ data. To do this we fit the mass spectrum
shown in Fig.~\ref{fig3} with a sum of signal and background distributions.
The background distribution shown in Fig.~\ref{fig3} is multiplied by a
scale factor $r_{\rm bkg}$, which is allowed to vary within a 12\%
uncertainty around unity. The signal cross section is described by 
Eq.~(\ref{signaleq}) with a constant nonresonant amplitude 
$P(E)=\sqrt{\sigma_{NR}}$ and the parameter $\sigma_R$ varied around the 
value listed in Table~\ref{tab0}.
From the fit we determine $\sigma_{NR}$ and $\varphi$. The result of the fit
is shown by the curve in Fig.~\ref{fig4} (right). The fitted value of the 
parameter 
$r_{\rm bkg}$ is 0.94. Therefore, the points in Fig.~\ref{fig4} (right) lie
slightly higher than those in Fig.~\ref{fig4} (left). The fit yields 
$\chi^2/\nu=11.0/12$ and the following values of parameters:
\begin{equation}
\sigma_{NR}=7.3^{+7.4}_{-5.3}\mbox{ pb},\;
\varphi=(-69\pm23)^\circ.\label{para1}
\end{equation}

We conclude that the \babar\ data on the $e^+e^-\to K_S K_L$ cross section do
not exclude the existence of the resonance with the parameters listed in Table~\ref{tab0},
but restrict the possible range of allowed values of the relative phase between
the resonant and nonresonant $e^+e^-\to K_S K_L$ amplitudes.

\section{\bf\boldmath
Simultaneous fit to the $e^+e^-\to K^+K^-$, 
$\pi^+\pi^-$, and $\pi^+\pi^-\eta$ data with an isovector
resonance\label{isovec}}
\begin{figure*}
\includegraphics[width=0.45\textwidth]{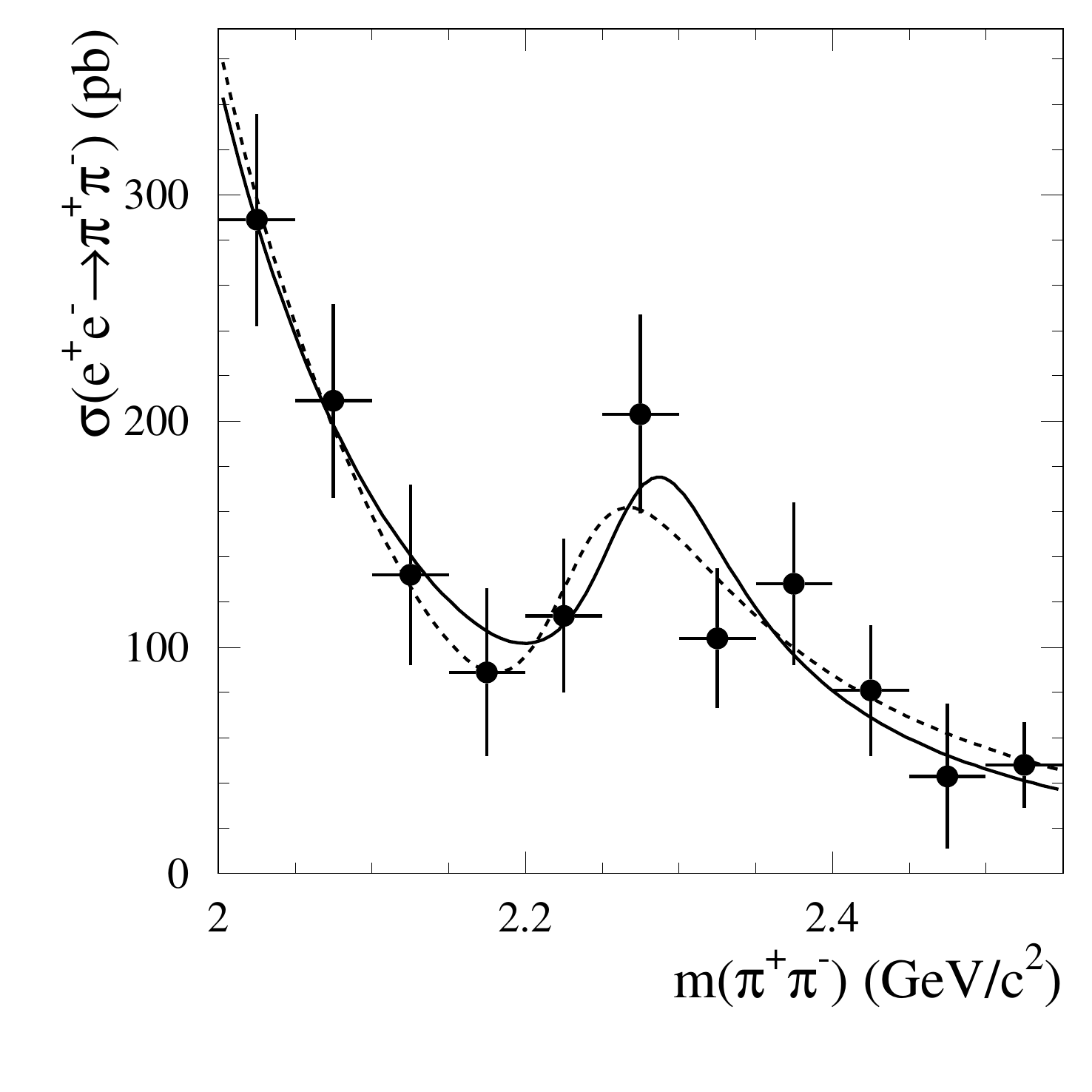}
\includegraphics[width=0.45\textwidth]{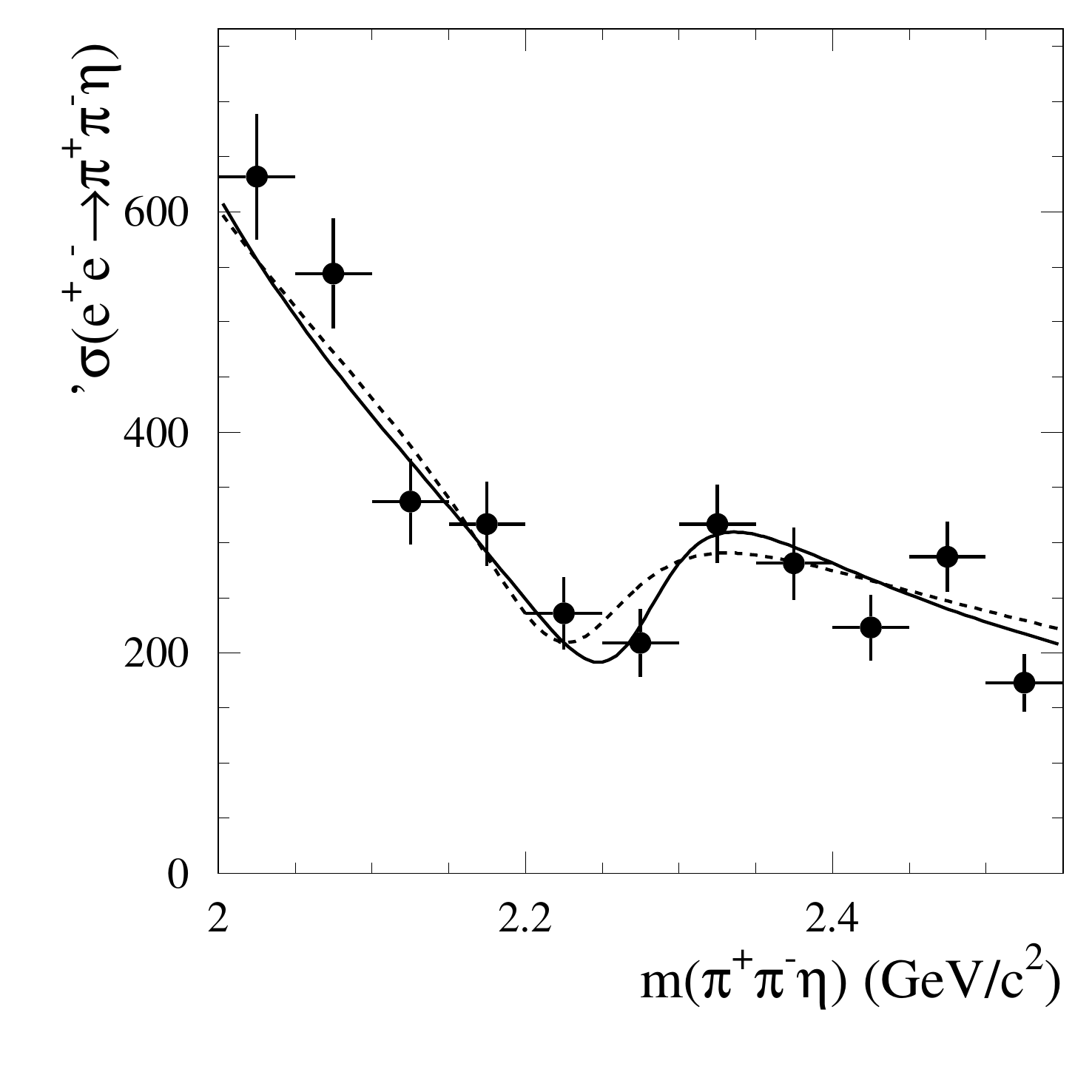}
\caption{Left panel: The $e^+e^-\to \pi^+\pi^-$ cross section measured by 
\babar~\cite{pipi}. Right panel: The $e^+e^-\to \pi^+\pi^-\eta$ cross section
measured by \babar~\cite{pipieta}. The solid curves are the results
of the simultaneous fit to the $e^+e^-\to \pi^+\pi^-$ and $\pi^+\pi^-\eta$ 
cross section data, while the dashed curves represent the results
of the simultaneous fit to the $e^+e^-\to K^+K^-$, $\pi^+\pi^-$, and 
$\pi^+\pi^-\eta$ cross section data.
\label{fig5}}
\end{figure*}
As discussed in the introduction, the mass and width of the resonance observed
in the process $e^+e^-\to K^+ K^-$ near 2.2~GeV are close to the parameters of
the state seen in the $e^+e^-\to \pi^+\pi^-$ cross section measured by
\babar~\cite{pipi}. 
The latter cross section in the energy range 2.00--2.55~GeV
is shown in Fig.~\ref{fig5} (left). An interference pattern in the energy
region near 2.25~GeV is also seen in the energy dependence of the 
$e^+e^-\to \pi^+\pi^-\eta$ cross section recently measured by 
\babar~\cite{pipieta} and shown in Fig.~\ref{fig5} (right). We perform a 
simultaneous fit to the $e^+e^-\to \pi^+\pi^-$ and $\pi^+\pi^-\eta$ data. The
cross sections are described by formulas similar to Eq.~(\ref{signaleq}). 
For the $\pi^+\pi^-\eta$ channel, the phase space factor 
$\beta(E)^3/\beta(M_{R})^3$ in Eq.~(\ref{signaleq}) is replaced by the factor
$p_\eta(E)^3/p_\eta(M_{R})^3 M_{R}/E$~\cite{achasov}, where $p_\eta$ is
the $\eta$-meson momentum calculated in the model of
 the $\rho(770)\eta$ intermediate state. The nonresonant
amplitude is described by the function $a/(E^2-b^2)$ inspired by the 
vector-meson dominance model. The ten fitted parameters are the mass ($M_R$)
and width ($\Gamma_{R}$) of the resonance, the peak cross sections
($\sigma(e^+e^-\to R\to \pi^+\pi^-)$ and $\sigma(e^+e^-\to R\to \pi^+\pi^-\eta)$),
and $a$, $b$, and $\varphi$ for the two channels. The result of the
fit is shown in Fig.~\ref{fig5} by the solid curves. The fit parameters
obtained are listed in the second column of Table~\ref{tab2}. The fit yields
$\chi^2/\nu=14.0/12$ ($P(\chi^2)=0.30$). The significance of the
resonance calculated from the difference in $\chi^2$ with and without the 
resonance contributions is 4.6$\sigma$. The systematic uncertainties in the
resonance parameters are determined as described in Sec.~\ref{fitkk}.
\begin{table*}
\caption{The parameters for the fit to 
the $e^+e^-\to \pi^+\pi^-$ and $\pi^+\pi^-\eta$ cross section data (second 
column), and to the $e^+e^-\to K^+K^-$, $\pi^+\pi^-$, and $\pi^+\pi^-\eta$
cross section data (third column). The quoted uncertainties are statistical 
and systematic, respectively.
\label{tab2}}
\begin{ruledtabular}
\begin{tabular}{ccc}
& $\pi^+\pi^-$ and $\pi^+\pi^-\eta$ & $K^+K^-$, $\pi^+\pi^-$, and $\pi^+\pi^-\eta$\\
\hline
$M_R$ (MeV/$c^2$)                  & $2270\pm 20\pm 9$ & $2232\pm 8\pm9$ \\
$\Gamma_{R}$ (MeV)                     & $116^{+90}_{-60}\pm50$ & $133\pm 14\pm 4$ \\
$\sigma(e^+e^-\to R\to K^+K^-)$ (pb) & $-$     & $41 \pm 6\pm 4$ \\
$\sigma(e^+e^-\to R\to \pi^+\pi^-)$ (pb) & $34^{+26}_{-19}\pm 4$ & $36^{+27}_{-20}\pm 4$ \\
$\sigma(e^+e^-\to R\to \pi^+\pi^-\eta)$ (pb) & $33^{+34}_{-13}\pm 4$ & $27^{+14}_{-11}\pm 4$ \\
$\varphi(e^+e^-\to  K^+K^-)$ (deg) & $-$     & $140 \pm 8\pm 9$ \\
$\varphi(e^+e^-\to  \pi^+\pi^-)$ (deg) & $147\pm30\pm 10$ & $188\pm19\pm9$ \\
$\varphi(e^+e^-\to  \pi^+\pi^-\eta)$ (deg) & $217\pm24\pm 9$ & $251\pm15\pm 9$ \\
$\chi^2/\nu$ &13.96/12& 17.2/14
\end{tabular}
\end{ruledtabular}
\end{table*}

We also perform a simultaneous fit to the BESIII and
\babar\ $e^+e^-\to K^+K^-$ data and the \babar\ $e^+e^-\to \pi^+\pi^-$ and
$\pi^+\pi^-\eta$ data. The $e^+e^-\to K^+K^-$ cross section
is parametrized as described in Sec.~\ref{fitkk}. The fit
parameters obtained are listed in the third column of Table~\ref{tab2}. Since
the $e^+e^-\to K^+K^-$ data are statistically more accurate than the
$\pi^+\pi^-$ or $\pi^+\pi^-\eta$ data, the fitted resonance
mass, width, and $\sigma(e^+e^-\to R\to K^+K^-)$ are similar to those 
(Table~\ref{tab0}) obtained in the fit to the $K^+K^-$ data alone. The results
of the fit for $e^+e^-\to \pi^+\pi^-$ and $\pi^+\pi^-\eta$ cross sections
are shown in Fig.~\ref{fig5} by the dashed curves. The $\chi^2/\nu$
calculated using the $\pi^+\pi^-$ and $\pi^+\pi^-\eta$ data is $17.2/14$ 
($P(\chi^2)=0.25$). We conclude that it is very likely that the interference
patterns observed in the three cross
sections discussed above are manifestations of the same isovector resonance,
$\rho(2230)$. It is interesting to note that the decay rates of this state
to $K^+K^-$, $\pi^+\pi^-$, and $\pi^+\pi^-\eta$ are all similar.. 

\section{Two-resonance fit}
\begin{figure}
\includegraphics[width=0.45\textwidth]{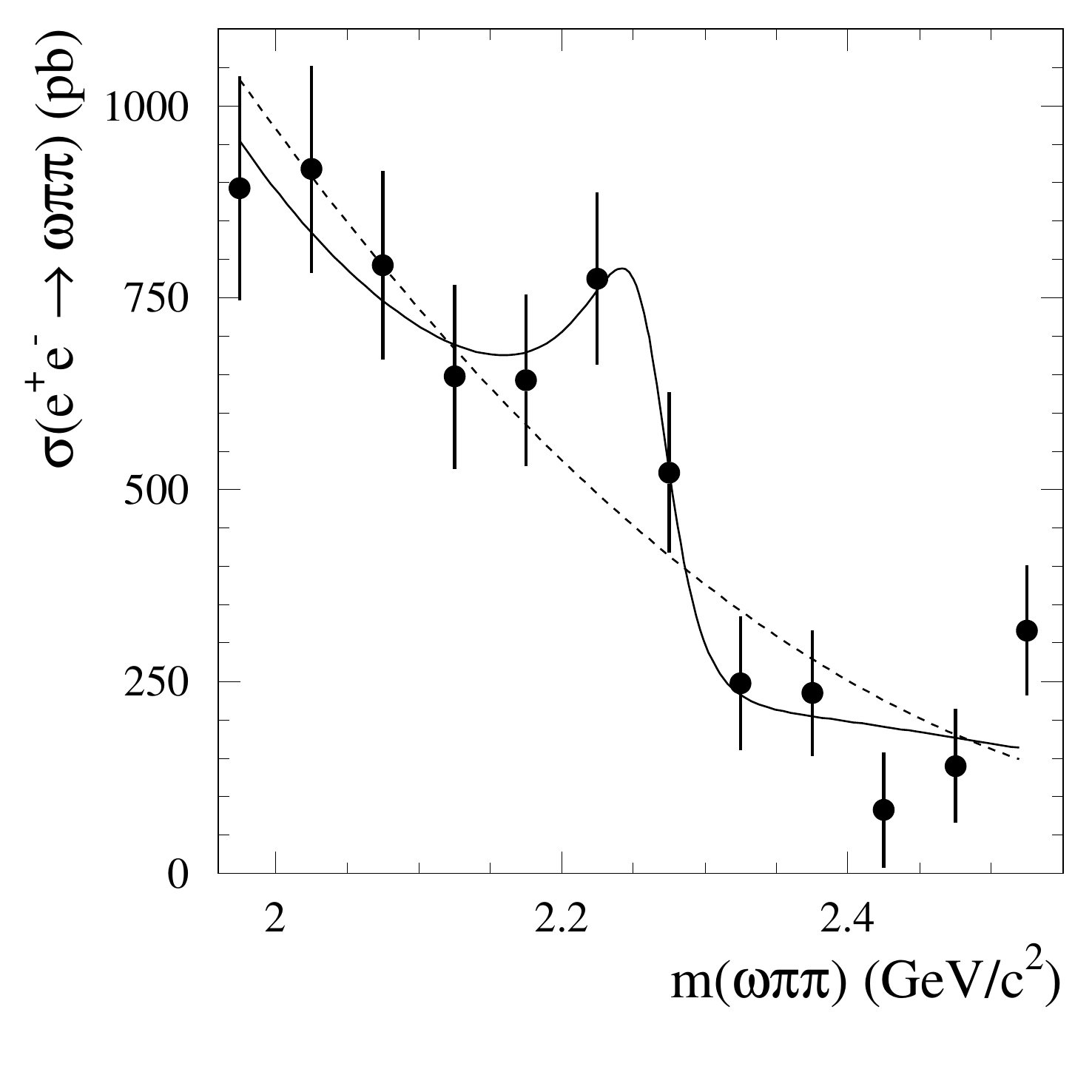}
\caption{The $e^+e^-\to \omega\pi\pi$ cross section measured by 
\babar~\cite{ompipi1,ompipi2}. The solid curve is the result of the fit by a 
coherent sum of resonant and nonresonant contributions, while the dashed 
curve represent the results of the fit to a quadratic polynomial.
\label{fig6}}
\end{figure}

The isovector state discussed in the previous section is expected to have an 
$\omega$-like isoscalar partner with a similar mass. An indication of an 
isoscalar resonance structure near 2.25~GeV is seen in the 
$e^+e^-\to \omega\pi^+\pi^-$ and $e^+e^-\to \omega\pi^0\pi^0$ cross sections
measured by \babar~\cite{ompipi1,ompipi2}. The energy dependence of the total 
$e^+e^-\to \omega\pi\pi$ ($\omega\pi^+\pi^- + \omega\pi^0\pi^0$) cross section
in the energy region of interest is shown in Fig.~\ref{fig6}. It is fitted
by a coherent sum of resonant and nonresonant contributions. We assume
that the process $e^+e^-\to \omega\pi\pi$ proceeds via the $\omega f_0(500)$
intermediate state. Therefore, the factor $\beta(E)^3/\beta(M_R)^3$ in
Eq.~(\ref{signaleq}) is replaced by the $s$-wave phase-space factor 
$p_\omega(E)/p_\omega(M_R)$, where $p_\omega$ is the $\omega$-meson momentum
in $e^+e^-\to \omega f_0(500)$. It should be noted that the phase-space factor
for the other possible intermediate state, $b_1(1235)\pi$, has a similar 
energy dependence in the energy region of interest.
The nonresonant amplitude is described by the function $a/(E^2-b^2)$.
The fit yields $\chi^2/\nu=6.8/6$. The result of the fit
is shown in Fig.~\ref{fig6} by the solid curve. The fitted resonance 
mass ($2265\pm20$~MeV/$c^2$) and width ($75^{+125}_{-27}$~MeV) are similar
to the parameters of the isovector state in Table~\ref{tab2}. 
Since different intermediate mechanisms (e.g., $\omega f_0$ and $b_1\pi$) 
contribute to the $\omega\pi\pi$ final state, the resonant and nonresonant
amplitudes may be not fully coherent. Inclusion in the fit of an incoherent 
contribution describing up to 50\% of the nonresonant cross section has an
insignificant impact on the fitted resonance mass and width. The dashed
curve in Fig.~\ref{fig6} is the result of the fit to data with a second-order
polynomial. The $\chi^2/\nu$ for this fit is 18.1/9. From the $\chi^2$ 
difference between the two fits we estimate that the significance of the resonance
signal in the $e^+e^-\to \omega\pi\pi$ cross section is $2.6\sigma$.

From isospin invariance, the isovector amplitude enters the 
$e^+e^-\to K^+K^-$ and $e^+e^-\to K_SK_L$ amplitudes with opposite sign
(in contrast to the isoscalar case)~\cite{kuhn}: 
\begin{eqnarray}
A(e^+e^-\to K^+K^-)&=&A_{I=0}+A_{I=1}, \nonumber\\
A(e^+e^-\to K_SK_L)&=&A_{I=0}-A_{I=1}.
\end{eqnarray}
The quark model predicts~\cite{kuhn} that the isoscalar amplitude related to
the $\omega$-like resonance is one-third the amplitude of the corresponding
$\rho$-like state and that these amplitudes have the same sign in the
$e^+e^-\to K^+K^-$ channel. If the $\rho$- and $\omega$-like resonances
have similar masses and widths, we expect the resonance amplitude in 
the $e^+e^-\to K_SK_L$ reaction to be about two times smaller than that in 
$e^+e^-\to K^+K^-$.  This weakens the constraints on the nonresonant
$e^+e^-\to K_SK_L$ cross sections and the interference phase,
relation (\ref{para1}),
obtained in the fit to the $e^+e^-\to K_SK_L$ data in Sec.~\ref{kskl}.
Repeating this fit with the resonance amplitude smaller by a factor of two,
we obtain $\chi^2/\nu=10.6/12$ and the parameters
\begin{equation}
\sigma_{NR}=5.0^{+8.2}_{-4.8},\;\varphi=(-51^{+56}_{-41})^\circ.\label{para2}
\end{equation}
The fit with zero nonresonant cross section also has an acceptable
$\chi^2$ value, $12.1/14$. We conclude that the two-resonance fit allows 
a simultaneous description of the $e^+e^-\to K^+K^-$ and $e^+e^-\to K_SK_L$
data without strong constraints on the interference parameters in 
the $e^+e^-\to K_SK_L$ channel.

Finally, we fit the $e^+e^-\to K^+K^-$, $e^+e^-\to \pi^+\pi^-$, and 
$e^+e^-\to\pi^+\pi^-\eta$ data using the model described in Sec.~\ref{isovec}
with an additional contribution from the $\phi(2170)$ resonance. The 
$\phi(2170)$ mass and width are fixed at their PDG values~\cite{pdg}. The 
inclusion of the $\phi(2170)$ has an insignificant impact on the quality of the
fit. The fitted value of the $\phi(2170)$ peak cross section is found to be 
consistent with zero, $0.8_{-0.8}^{+2.9}$ pb. 

\section{Summary}
In this paper, we present measurements of the $e^+e^-\to K_SK_L$ cross section
in the center-of-mass range from 1.98 to 2.54 GeV.
The measured cross section is consistent with zero and does not exhibit
evidence for a resonance structure.
The $K_SK_L$ data are analyzed in conjunction with BESIII~\cite{beskkc}
and \babar~\cite{babarkkc} data on the $e^+e^-\to K^+K^-$ cross section,
and with \babar\ data on the $e^+e^-\to \pi^+\pi^-$~\cite{pipi},
$\pi^+\pi^-\eta$~\cite{pipieta},
$\omega\pi^+\pi^- + \omega\pi^0\pi^0$~\cite{ompipi1,ompipi2} cross sections to
examine properties and better elucidate the nature of the resonance structure
observed by BESIII in the $e^+e^-\to K^+K^-$ cross section near 
2.25~GeV~\cite{beskkc}.

The interference patterns seen in the $e^+e^-\to \pi^+\pi^-$
and $e^+e^-\to\pi^+\pi^-\eta$ data near 2.25~GeV provide $4.6\sigma$ evidence
for the existence of the isovector resonance $\rho(2230)$. Its mass and width 
are consistent with the parameters of the resonance observed in the 
$e^+e^-\to K^+K^-$ channel. All three cross sections are well   
described by a model with $\rho(2230)$ mass and width  
$M=2232\pm 8\pm9$~MeV/$c^2$ and $\Gamma=133\pm 14\pm4$~MeV. 

Any resonance in the $e^+e^-\to K^+ K^-$ cross section should also be manifest
in the $e^+e^-\to K_S K_L$ cross section.
The \babar\ data on the $e^+e^-\to K_S K_L$ cross section do not exclude the 
existence of the $\rho(2230)$ resonance, but strongly restrict the possible 
range of allowed values of the relative phase between
the resonant and nonresonant $e^+e^-\to K_S K_L$ amplitudes.
This restriction may be significantly weakened by inclusion in the fit of an
additional isoscalar resonance with a nearby mass. An indication of
such a resonance with $2.6\sigma$ significance
is seen in the $e^+e^-\to \omega\pi\pi$ cross section.

Further study of the resonance structures near 2.25~GeV can be performed
at the BESIII experiment, where the cross sections for 
$e^+e^-\to\pi^+\pi^-\eta$, $\omega\pi^+\pi^-$, $\omega\pi^0\pi^0$ and
other exclusive processes in the energy range between 2 and 2.5 GeV may be 
measured with high accuracy.

\begin{acknowledgments}
We are grateful for the 
extraordinary contributions of our PEP-II colleagues in
achieving the excellent luminosity and machine conditions
that have made this work possible.
The success of this project also relies critically on the 
expertise and dedication of the computing organizations that 
support \babar.
The collaborating institutions wish to thank 
SLAC for its support and the kind hospitality extended to them. 
This work is supported by the
US Department of Energy
and National Science Foundation, the
Natural Sciences and Engineering Research Council (Canada),
the Commissariat \`a l'Energie Atomique and
Institut National de Physique Nucl\'eaire et de Physique des Particules
(France), the
Bundesministerium f\"ur Bildung und Forschung and
Deutsche Forschungsgemeinschaft
(Germany), the
Istituto Nazionale di Fisica Nucleare (Italy),
the Foundation for Fundamental Research on Matter (The Netherlands),
the Research Council of Norway, the
Ministry of Education and Science of the Russian Federation, 
Ministerio de Econom\'{\i}a y Competitividad (Spain), the
Science and Technology Facilities Council (United Kingdom),
and the Binational Science Foundation (U.S.-Israel).
Individuals have received support from 
the Marie-Curie IEF program (European Union) and the A. P. Sloan Foundation (USA). 
\end{acknowledgments}


\end{document}